\documentclass[showpacs,superscriptaddress,11pt]{revtex4}
\usepackage{doi}
\usepackage{hyperref}
\hypersetup{
  colorlinks=true,        
  linkcolor=blue,         
  citecolor=cyan,         
}

\usepackage{graphicx}
\usepackage{dcolumn}
\usepackage{bm}
\usepackage{color}
\usepackage{enumitem}
\usepackage{amsmath}
\usepackage{amssymb}

\begin{document}

\title{Weak gravitational lensing and shadow of a GUP-modified Schwarzschild black hole in the presence of plasma}

\author{Husanboy Hoshimov}

\affiliation{Fergana State University, Murabbiylar St 19, Fergana 150100, Uzbekistan}

\author{Odil Yunusov}

\affiliation{Ulugh Beg Astronomical Institute, Astronomy St 33, Tashkent 100052, Uzbekistan}
\affiliation{Institute of Theoretical Physics, National University of Uzbekistan, Tashkent 100174, Uzbekistan}

\author{Farruh Atamurotov}

\affiliation{New Uzbekistan University, Movarounnahr street 1, Tashkent 100000, Uzbekistan}
\affiliation{Central Asian University, Milliy Bog' Street 264, Tashkent 111221, Uzbekistan}
\affiliation{Institute of Fundamental and Applied Research, National Research University TIIAME, Kori Niyoziy 39, Tashkent 100000, Uzbekistan}
\affiliation{Tashkent State Technical University, Tashkent 100095, Uzbekistan}

\author{Mubasher Jamil}
\email{mjamil@sns.nust.edu.pk (corresponding author)}
\affiliation{School of Natural Sciences, National University of Sciences and Technology, Islamabad, 44000, Pakistan}

\author{Ahmadjon Abdujabbarov}

\affiliation{Ulugh Beg Astronomical Institute, Astronomy St 33, Tashkent 100052, Uzbekistan}
\affiliation{Institute of Theoretical Physics, National University of Uzbekistan, Tashkent 100174, Uzbekistan} 
\affiliation{University of Tashkent for Applied Sciences, Str. Gavhar 1, Tashkent 100149, Uzbekistan}

\date{\today}

\begin{abstract}

In this work, we have studied weak gravitational lensing effect around black hole and determine shadow radius in GUP-corrected-Schwarzschild spacetime (S-GUP) in presence of plasma environment. We started with orbits of photons around black hole in S-GUP. In addition, we have studied shadow and gravitational weak lensing around such black hole. By using observational data of EHT project for the M87* and Sgr A*, we have got constrains parameter $\epsilon$ in S-GUP gravity. Further to make a connection with observations, we investigate magnification and  position of images formed as a result of lensing, and finally weak deflection angle and magnification for sources near M87* and Sgr A*.

\end{abstract}
\maketitle
\newpage
\section{Introduction} 

The problem of unifying quantum theory and gravity remains one of the most intriguing and formidable challenges in contemporary physics. At the very center of this quest lies the attempt to reconcile two of the most successful yet fundamentally different theories: quantum mechanics, which governs the behavior of particles at the subatomic scale, and general relativity, which describes the curvature of spacetime due to mass and energy.
While quantum mechanics has been remarkably successful in explaining phenomena on atomic and subatomic scales, the general relativity has provided an exquisite framework for understanding gravity's effects from solar to cosmic scales.
 Attempts to merge these theories, like string theory, loop quantum gravity, and others, have illuminated the complexity of the task. Fortunately, black holes are not only gravitational but also quantum mechanical objects and hence work as a laboratory for testing predictions of quantum gravity as well as provide a theoretical framework for pursuing the goal of theory of quantum gravity \cite{Susskind:1995qc}. 

The generalized uncertainty principle (GUP) is employed as a useful tool against the shortcomings of general relativity (GR). The uncertainty relationship between position and momentum operators due to GUP incorporates the nonlinear terms due to uncertainties in momentum $([\hat{x},\hat{p}]=\iota \hbar (1+\beta \hat{p}^2))$ which are motivated both from string theory and loop quantum gravity \cite{Ali:2009zq}. Since GUP implies the minimal length at Planck scale, it removes the singularity predicted by standard GR. Since then, physicists have studied the GUP extensively and profoundly \cite{Nasser2015RPPh}. The GUP has been considered various ways, such as based on string theory \cite{Veneziano1986EL,Amati1987PhLB,Amati1989,Gross1987PhLB,Gross1988NuPhB}. In the litarature, have been considered various tests of the GUP corrected BH \cite{Adler2001GReGr,2020EL,Karmakar2022IJMPA,Fu2021NuPhB,Li2017GReGr,Rizwan2017IJMPD}.
The effect of GUP on the accretion disk onto Schwarzschild black hole is considered Ref. \cite{Moussa:2023},we find the quantum correction of the Schwarzschild black  hole metric based on GUP \cite{Lenat83a}. Additionally, in the literature, GUP parameter estimates various physical effects, for example, constraint by gravitational wave events \cite{Feng2017PhLB}, through computing perihelion precession, both for planets in the solar system and for binary pulsars \cite{Scardigli2015EPJC}, shadow \cite{Neves2020EPJC}, from Shapiro time delay, gravitational redshift, and geodetic precession for the GUP modified Schwarzschild metric \cite{Okcu:2021oke}. Further, the GUP corrected Kerr black hole is studied and constrained via quasi-periodic oscillations and S2 star motion about the Milky Way galactic center \cite{Jusufi:2020wmp}. Also, the esoteric objects like wormholes have been constructed via Casimir energy density with GUP corrections \cite{Jusufi:2020rpw}.

Plasma effects are vital in the study of the properties of the GUP-modified Schwarzschild black hole, influencing weak gravitational lensing and shadow features. The plasma introduces modifications in the spacetime geometry, affecting the trajectory of light rays and altering the observable characteristics. Understanding these effects is crucial for accurately interpreting astrophysical observations, providing insights into the complex interplay between quantum gravity corrections and the surrounding plasma environment~\cite{Bisnovatyi2010,Rog:2015a,43,Babar21a}.
The imperative of rigorously testing gravity models through diverse observational techniques, such as weak lensing and particle motion, is paramount to deriving meaningful constraints on these models~\cite{patla2008AJ}. 
Weak lensing, by scrutinizing the bending of light due to gravitational fields, offers insights into the distribution of matter and energy across vast cosmic scales~\cite{Synge1960,Azreg-Ainou:2020bfl}. By comparing observational outcomes with predictions from gravity models, we can discern whether a particular model accurately describes the universe's behavior. Particularly, the recent observation of gravitational waves within LIGO-Virgo collaboration~\cite{Ligo2016} and image of supermassive black hole M87*~\cite{2} and Sgr A*~\cite{Akiyama2022sgr} within Event Horizon Telescope collaboration open new window to construct new tests of modified and alternative theories of garvity~\cite{Sunny2023cons,kalita2023,Ulislam2023,Reggie2023aaa,Walia2022aaa,2023ChPhC..47b5102A,Farruh2022cosn,Nampalliwar:2021tyz}.  By observing the M87 galactic center at the 230 GHz, the EHT probe imaged the polarized emission around supermassive black hole at the scale of the event horizon. The EHT team also found plasma and magnetic field structure about supermassive black hole in M87 galaxy which lead to determine the accretion rate of the central black hole as well \cite{EventHorizonTelescope:2021srq}. The EHT team also determined the electron temperature close to $10^{10}$ Kelvin and the average number density of electron $10^{4-7}cm^{-3}$. Further, the accretion rate of M87 central supermassive black hole is found to be $(3-20)\times 10^{-4}$ solar mass per year. This is very first observational result of its kind which confirms the presence of plasma environment near a supermassive black hole. Although the EHT results clearly demonstrate the presence of a magnetized plasma near the supermassive black hole, there is still not available a better theory to model such magnetized plasma near a black hole. Therefore we focus on the non-magnetized plasma near the black hole.

Note that the plasma is a highly refractive and dispersive medium so that the light passing through the plasma medium changes its path by interacting with the plasma. The deflection angle of light is affected not only by the gravitational field but also due to plasma effects. The plasma model near a black hole is modelled as a static medium and spatially varying density profile which falls radially outside the black hole \cite{Li:2021btf}. The plasma particles can be dragged closed to the black hole due to gravitational effects, as close as the innermost stable circular orbit, and lead to plasma accretion by the black hole as well as formation of jets \cite{Chou:1999qna}. Hence the deflection angle calculations of light rays emitted near the black hole horizon must take into account the plasma effects. Further, from analytical perspective, the overall shape of the black hole shadow is either bigger or smaller compared to black hole shadow without plasma as it depends on several plasma parameters such as plasma density \cite{Chowdhuri:2020ipb}.   The test particle and photon motion, serves as a fundamental probe of gravity's behavior under different conditions. More recently, the EHT team has observed photon rings around the supermassive black hole in M87* galactic center and estimated the black hole mass nearly 7 billion solar mass \cite{Broderick:2022tfu}. Deviations from the predictions of established theories could signal the presence of novel gravitational effects.
On the other hand the plasma environment around black hole sufficiently affects on the trajectory of photons and cannot be neglected~\cite{Bisnovatyi2010,Perlick2015PhRvD,Perlick17a,Rog:2015a,Atamurotov15a,Babar20,Atamurotov21axion,2023EPJC...83..318G}. Furthermore, different types of plasma distributions around black holes within other gravity models have been extensively studied in~\cite{Tsp:2009a,45,43,Babar21a,49,Babar2021b,12,Car:2018a,47,Turi:2019a,Far:2021a,Atamurotov2022EPJP,2023Symm...15..848A,Atamurotov2024NewA,Orzuev2024NewA,2023ChPhC..47c5106A}. 

The integration of above mentioned tests contributes to refining our understanding of gravity's intricacies and its role in the universe's evolution. Ultimately, these observations play an indispensable role in constraining and validating gravity models, steering us toward a more comprehensive and accurate portrayal of the fundamental forces that shape our cosmos. With this aim, here we plan to study the photon motion and effect of weak gravitational lensing around black hole within GUP modified gravity model.  
The paper is organized as follows. In Sect.~\ref{null}, we provide null geodesic in presence of plasma. In Sect.~\ref{shadow}, we probe the GUP modified Schwarzschild metric with BHs shadow. In Sect.~\ref{weak lensing}, we calculate deflection angle and magnfication in weak field limit. Finally, in Sect.~\ref{conlusion}, we summarize obtained results.

\section{Null Geodesic in Gup-Modified schwarzschild spacetime} \label{null}

In Boyer-Lindquist coordinates, S-GUP metric is given by \cite{Scardigli2015EPJC}
\begin{align}
    ds^2=-f(r) d t^2+f(r)^{-1} d r^2+r^2(d \theta^2 +\sin^2 \theta d\phi^2 ) ,
    \label{metric}
\end{align}
with 
\begin{align}
   f(r) = 1-\frac{2 M}{r}+\epsilon\frac{M^2}{r^2} ,
\end{align}
where $\epsilon$ is a dimensionless parameter representing the GUP correction \cite{Scardigli2015EPJC} which has been constrained $ -44.9<\epsilon\le 1$ \cite{Okcu:2021oke}. In addition, this metric reduces to the Schwarzschild metric when $\epsilon \to 0$.

For a photon moving in a plasma, the Hamiltonian is written as  \cite{Synge1960} 
\begin{equation}\label{hamil}
     H(x^\alpha, p_\alpha)=\frac{1}{2}\tilde{g}^{\alpha \beta}p_\alpha p_\beta,
\end{equation}
here $x^\alpha$ describe the spacetime coordinates and $\Tilde{g}_{\alpha\beta}$ is the effective metric which can be written as the form
\begin{align}
    \tilde{g}^{\alpha \beta}=g^{\alpha \beta}-(n^2-1)u^\alpha u^\beta,
\end{align}
where $n$ is the refractive index of the plasma, $p_\alpha$ and $u^\beta$ denote four-momentum and four-velocity of the photon, respectively. The refraction index $n$ of this plasma is \cite{Rodrigues2020PhRvD}:
\begin{equation}\label{refr}
    n^2=1-\frac{\omega{_p}(r)^2}{\omega(r)^2}
\end{equation}
where $\omega^2_p(r)=4 \pi e^2 N (r)/m_e$ ($e$ and $m_e$ are the electron charge and mass, respectively, and $N$ is the number density of the electrons) is electron plasma frequency and the photon frequency $\omega(r)$ measured by a static observer is found by using the gravitational redshift formula
   \begin{equation}
       \omega(r)=\frac{\omega_0}{\sqrt{f(r)}}.
   \end{equation}
Here $\omega_0=\rm{const}$ is the frequency at infinity ($f(\infty)=1$) and $\omega_0=\omega(\infty)=-p_t$ , which denote the energy of the photon at spatial infinity \cite{Perlick2015PhRvD}. Light can propagate in the plasma only if its frequency is larger than the plasma frequency, hence Eq. (\ref{refr}) is valid only when $\omega_p/\omega<1$, otherwise the photon does not propagate in plasma medium. If $\omega_0\simeq\omega$, deflection angle is much larger than the vacuum case ($\omega_p=0$), $\alpha\gg2R/b$, where $b$ is the impact parameter. 


By using Eqs.~(\ref{hamil}) and (\ref{refr}) and $x^{\alpha}=\delta H/\delta p_{\alpha}$ relationship the components of the four-velocity vector for photons in the equatorial plane $(\theta=\frac{\pi}{2},p_\theta=0)$ can be written as follows 

    \begin{align}
        &\dot{t}\equiv\frac{dt}{d\lambda}=\frac{-p_t}{f(r)},\label{tdot}\\ 
        &\dot{r}\equiv\frac{dr}{d\lambda}=p_r f(r) ,\label{rdot}\\ 
        &\dot{\phi}\equiv\frac{d\phi}{d\lambda}=\frac{p_\phi}{r^2}.\label{phidot}
    \end{align}
 From Eqs (\ref{rdot}) and (\ref{phidot}), we have an equation for the phase trajectory of light 
\begin{equation}
    \frac{dr}{d\phi}=\frac{\dot{r}}{\dot{\phi}}=\frac{f(r)r^2 p_r}{p_\phi}.
\end{equation}
Using the constraint $H=0$ for the photon motion, we can rewrite the above equation as
\begin{equation}
    \frac{dr}{d\phi}=\pm r\sqrt{f(r)}\sqrt{h^2(r)\frac{\omega^2_0}{p^2_\phi}-1},
\end{equation}
where we define \cite{Perlick2015PhRvD} 
\begin{equation}\label{define}
    h^2(r)\equiv r^2\Bigg[\frac{1}{f(r)}-\frac{\omega^2_p(r)}{\omega^2_0}\Bigg].
\end{equation} 
The radius of a circular orbit of light around black hole, particularly the one that forms a photon sphere of radius $r_p$, is determined by solving the following equation as \cite{Perlick:2021aok}
\begin{equation}\label{for:rph}
    \frac{d(h^2(r))}{dr}\Big|_{r=r_p}=0
\end{equation}
After substitung Eq.(\ref{define}) into Eq.(\ref{for:rph}) one can write the algebraic equation for radius of photon $r_p$ in the presence of a plasma medium as
\begin{equation}
    \frac{r^2_p(-3Mr_p+r_p^2+2\epsilon M^2)}{(-2Mr_p+r_p^2+\epsilon M^2)^2}=\frac{\omega^2_p(r_p)}{\omega_0^2}+\frac{r_p\omega_p(r_p)\omega'_p(r_p)}{\omega_0^2},
    \label{photon-r}
\end{equation}
where prime denotes the derivative with respect to radial coordinate $r$. Obviously, the roots of the equation cannot be obtained
analytically for most variants of  $\omega_p(r)$; however we will consider a few simplifed cases below.

\subsection{Homegeneous plasma with $\omega_p^2(r)=const.$}

\begin{figure}
    \centering
    \includegraphics[scale=0.6]{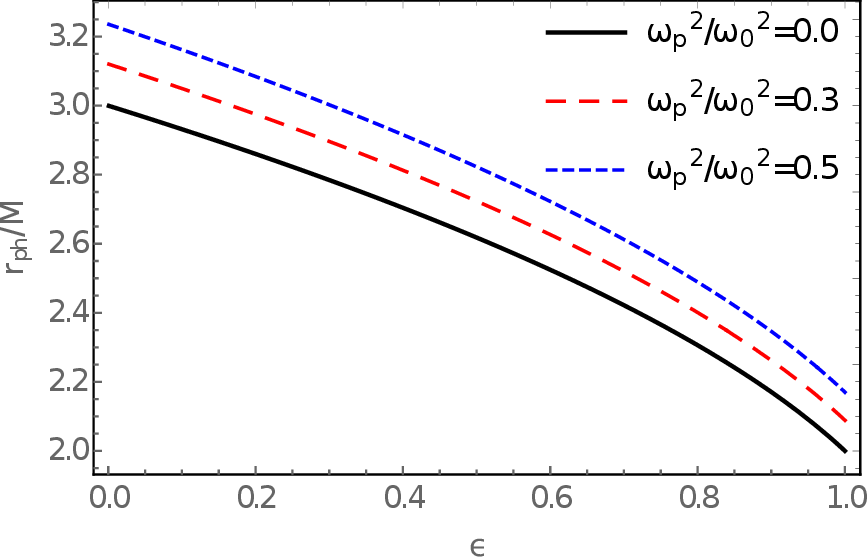}
    \caption{This graph describes radius of the photon sphere in a homogeneous plasma medium with constant frequency $\omega_p=\rm{const}$.}
    \label{photon1}
\end{figure}

First of all, in the case of homogeneous plasma medium with a constant plasma frequency throughout the medium Eq~(\ref{photon-r}) can be solved numerically and it is illustrated in Fig~(\ref{photon1}). From this picture one can easily see that the photon's radius around black hole decreases with increasing GUP parameters $\epsilon$. As well as, the plasma medium increases radius of the photon sphere.

\subsection{Inhomogeneous plasma with $\omega^2_p(r)=z_0/r^q$}
\begin{figure}
    \centering
    \includegraphics[scale=0.55]{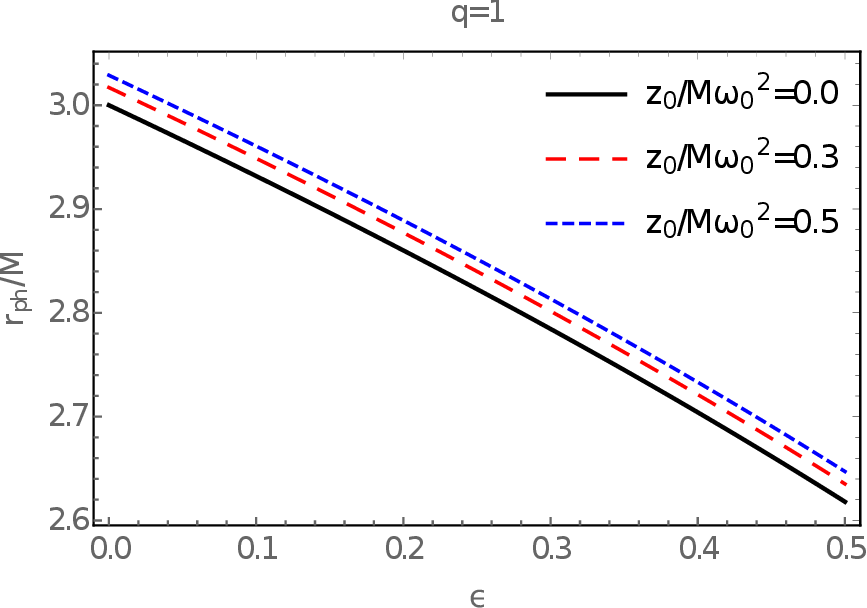}
    \includegraphics[scale=0.55]{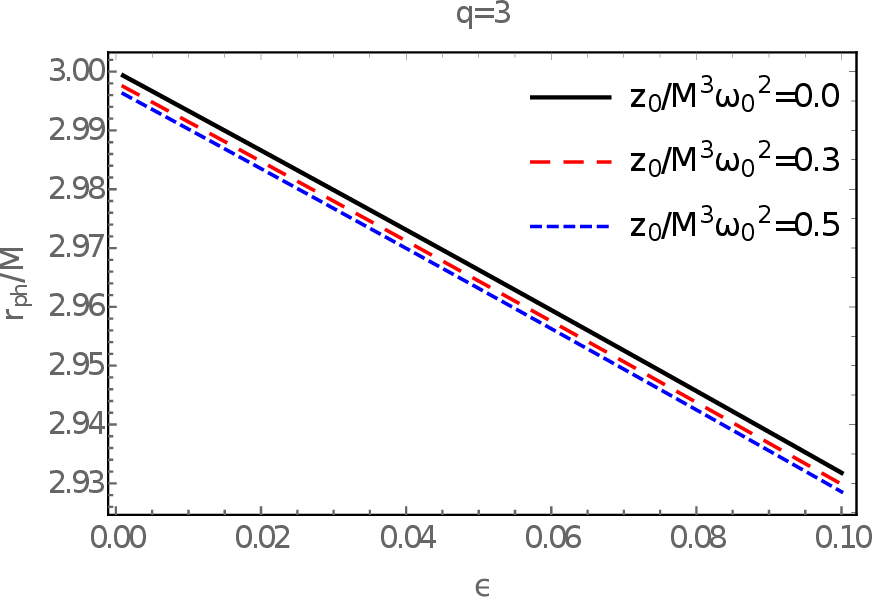}
    \caption{This graph shows a variation of $ \ r_ {ph}/M$ with respect to $\epsilon $ for the case $q=1$ (left panel) and $q=3$ (right panel).}
    \label{fig2}
\end{figure}

In this subsection we probe photon spheres in the presence of an non-uniform plasma, where the plasma frequency must satisfy a simple power-law of the from \cite{Rog:2015a,Xinzhong2018MNRAS},
\begin{align}\label{inhomogenius}
        \omega_p^2(r)=\frac{z_0}{r^q},
\end{align}
where $z_0$ and $q>0$ are free parameters. To analyze the primary features of the power-law model, we restrict ourselves to the following cases: $q=1$ and $z_0$ as a constant which reproduces the negativ-mass diverging lens exactly\cite{Xinzhong2018MNRAS} and $q=3$ and $z_0$ as a constant which is related to the stellar surface based on Goldreich-Julian(GJ) density. Using Eqs.(\ref{photon-r}), (\ref{inhomogenius}) we obtain the radius of the photon sphere based on numerical scheme for the inhomegeneous plasma medium, as illustrated in Fig.(\ref{fig2}). This profile shows that the radius of photon sphere decreases when $\epsilon$ parameter increases. On the other hand, plasma medium leads to the widening of the photon sphere radius if its distribution obeys $\omega_p^2(r)=z_0/r$ law (shown in left panel), contrarily, in the case of $q=3$, the effect of plasma is negative i.e. the presence of plasma around black hole  slightly shrinks the photon radius (shown in right panel). Furthermore, difference of the photon radius in the case of without and with plasma is enough small. This suggests that testing and distinguishing the homogeneous plasma from inhomogeneous plasma around BHs based on their shadows may be quite challenging. 

\section{Shadow of a BH surrounded by plasma medium}\label{shadow}

We investigate the radius  of the shadow of S-GUP spacetime metric in a plasma medium. The angular radius of the BH shadow, $\alpha_{sh}$, is defined by geometric approach, which results in the following expression \cite{Perlick2015PhRvD,Atamurotov21axion}
\begin{equation}
    \sin^2{\alpha_{sh}}=\frac{ h^2(r_p)}{h^2(r_0)}=\frac{r^2[\frac{1}{f(r_p)}-\frac{\omega_p^2(r_p)}{\omega_0^2}]}{r_0^2[\frac{1}{f(r_0)}-\frac{\omega^2_p(r_0)}{\omega_0^2}]},
\end{equation}
where $r_0$ and $r_p$ represent the locations of the observer and the photon sphere, respectively. Notably, if the observer is located at a sufficiently large distance from the BH, one can approximate the radius of the BH  shadow as follows \cite{Perlick2015PhRvD} 
\begin{equation}
    R_{sh} \approx r_0 \sin{\alpha_{sh}} = \sqrt{r^2_p\Bigg[\frac{1}{f(r_p)}-\frac{\omega_p^2(r_p)}{\omega_0^2}\Bigg]}.
\end{equation}

\begin{figure}
    \centering
    \includegraphics[scale=0.7]{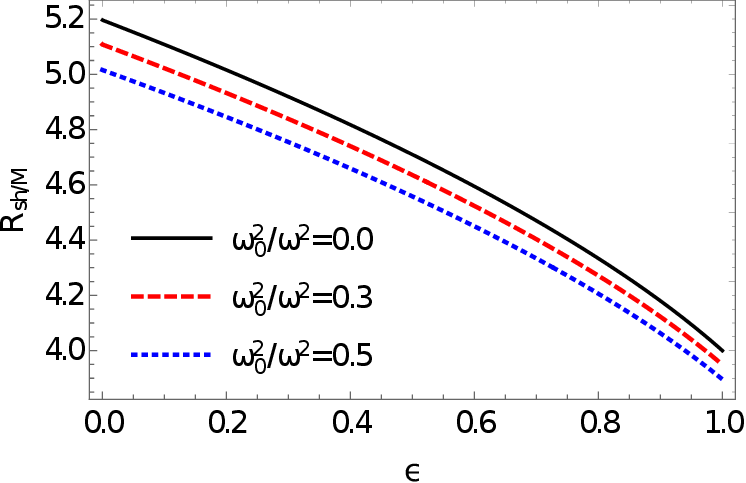}
    \caption{This graph describes a variation of $R_{sh}/M$ with respect to $\epsilon$ in uniform plasma}
    \label{Shadow uniform}
\end{figure}

This is based on the fact that $h(r) \to r$, which follows from Eq.(\ref{define}) at spatial infinity for both the models of plasma.For vacuum $\omega(r)\equiv0,$ we recover the radius of the Schwarzshild BH shadow, $ R_{sh}=3\sqrt{3}M$, when $r_p=3M.$ The radius of the BH shadow is depicted for different parameter $\epsilon$ in Fig~(\ref{Shadow uniform}) for a homogeneous plasma frequency. This figure illustrates that the shadow  radius decreases much more steeply with $\epsilon$ as well as plasma frequency. Furtermore, we have investigated inhomogeneous plasma with $q=1$ and $q=3$ in Eq.(\ref{inhomogenius}). One can see that due to the presence of plasma radius of the BH's shadow is decreasing. 

\begin{figure}
    \centering
    \includegraphics[scale=0.55]{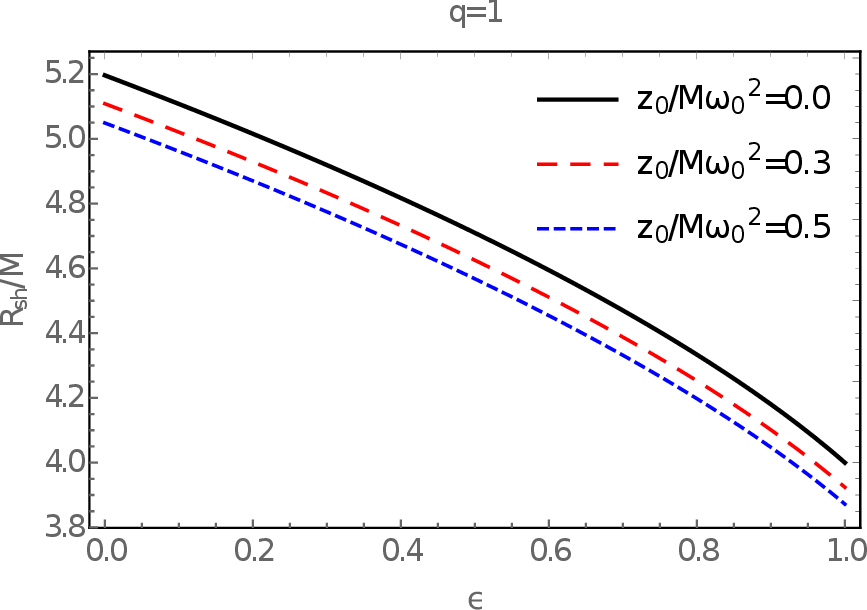}
    \includegraphics[scale=0.55]{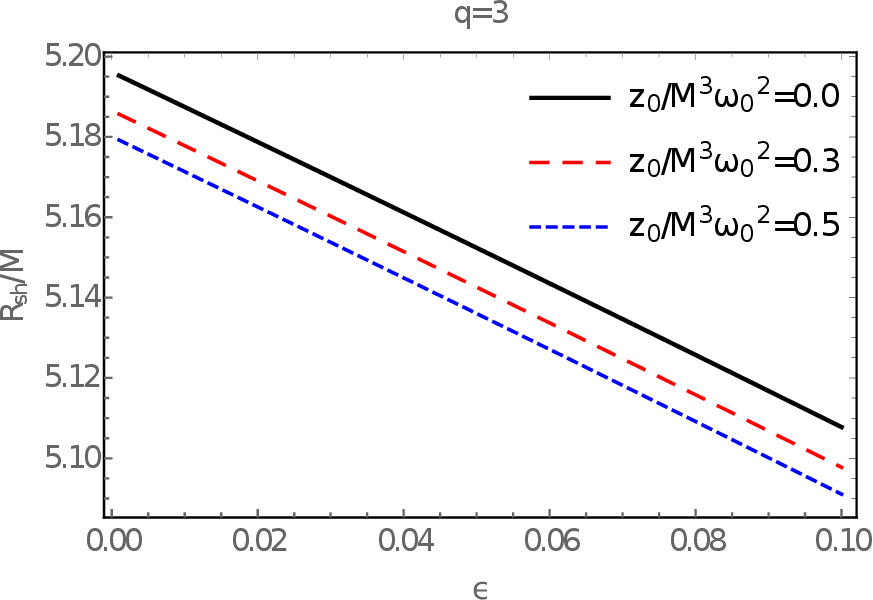}
    \caption{Variation of $ R_{sh}/M $ with respect to $\epsilon$ for $q=1$ (left panel) and $q=3$ (right panel).} 
    \label{Nonuniform Shadow}
\end{figure}

\begin{figure}
    \centering
    \includegraphics[width=0.45\textwidth]{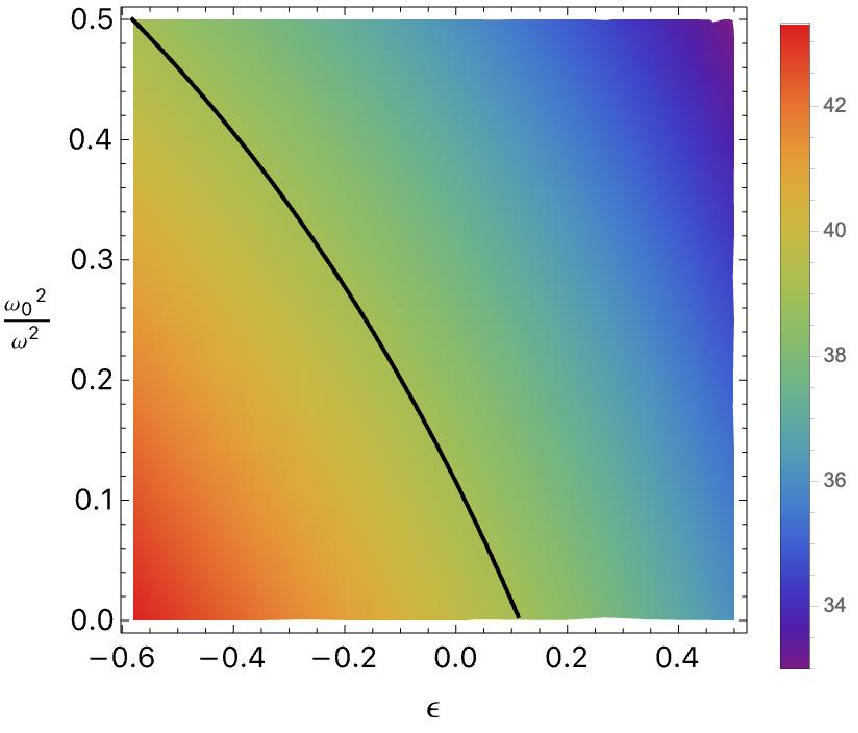}
    \includegraphics[width=0.45\textwidth]{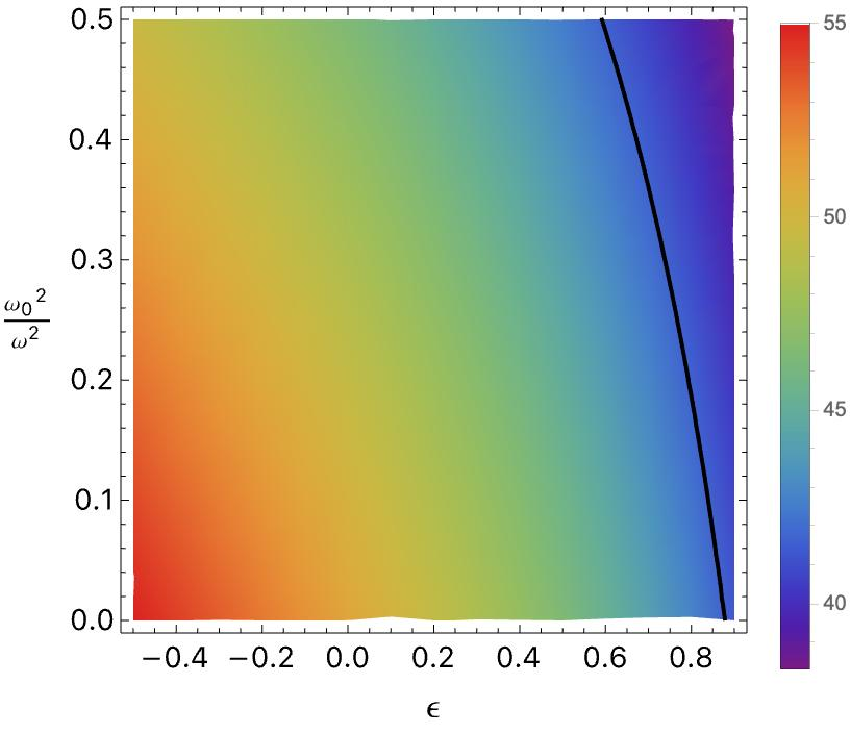}
    \caption{Relationship between $\epsilon$ and plasma parameter via observational shadow angular diameter of M87* (left panel) and Sgr A* (right panel) black holes.  Black lines correspond to black hole shadow angular diameter $\theta_{sh}=39\mu as$ for M87* (left panel) and $\theta_{sh}=41.7\mu as$ for Sgr A* (right panel), respectively. Regions contained within it satisfy M87*(left panel) and Sgr A* (right panel) black holes shadow 1-$\sigma$ bound. }
    \label{Constraint}
\end{figure}

Now we assume that the supermassive BHs M87* and Sgr A* are spherically symmetric static. Although the observation got by the EHT collaboration does not support the assumption taken here. However, here we explore theoretically the constrain on the parameter $\epsilon$, from the data provided by the EHT project. To constrain this parameter $\epsilon$ we use the observational data released by the EHT project for the BH shadows of the supermassive BHs M87* and Sgr A*. The angular diameter of the shadow, the distance from sun system and the mass of the BH at the centre of the galaxy M87, are $\Omega_\text{M87*} = 42 \pm 3 \:\mu$as,  $D = 16.8 \pm 0.8$ Mpc and $M_\text{M87*} = (6.5 \pm 0.7)$x$10^9 \: M_\odot$, respectively \cite{2}. For the Sgr A* the data recently obtained by the EHT project is  $\Omega_\text{Sgr A*} = 48.7 \pm 7 \:\mu$as, $D = 8277\pm9\pm33$ pc and $M_\text{Sgr A*} = 4.297 \pm 0.013$x$10^6 \: M_\odot$ (VLTI) \cite{Akiyama2022sgr}. Using this data, we can estimate the diameter of the shadow cast by the BH, per unit mass from the following expression \cite{Sunny2023cons},
\begin{equation}
    d_\text{sh} = \frac{D \Omega}{M}\,.
    \label{43}
\end{equation}
Now we can obtain the diameter of the shadow from the expression $d_\text{sh}^\text{theo} = 2R_\text{sh}$. Thus, the diameter of the BH shadow image is 
$d^\text{M87*}_\text{sh} = (11 \pm 1.5)M$ for M87* and  $d^\text{Sgr A*}_\text{sh} = (9.5 \pm cc-by)M$ for Sgr A*. From the data by the EHT collaboration, we obtain the constrain on the parameter $\epsilon$ for the supermassive BHs at the centre of the galaxy M87 and the Sgr A. We present our results obtained here in the Fig.~\ref{Constraint}. In this figure, it is observed that the angular diameter of BH shadow decreases with the increasing value of parameter $\epsilon$. It is also observed that the angular diameter of shadow for M87* and Sgr A* BH in the context of the S-GUP black hole is smaller than the other ordinary astrophysical BH such as Schwarzschild BH.
In addition, by using observational data of EHT project, we can get constrains parameter $\epsilon$. Fig.(\ref{Constraint}) shows that the region that is coincided observation. As a result, we can get upper bound of parameter $\epsilon$ of GUP-modified Schwarzschild black hole. This value is $\epsilon_{\rm upper}=0.2$ in the case of M87*, while it can be $\epsilon_{\rm upper}=0.9$ in the case of SgrA*.

\section{Gravitational weak lensing in presence of the plasma medium}\label{weak lensing}

In this section our main concern is to unravel the effects of gravitational lensing in the Schwarzchild black hole with GUP surrounded by a plasma considering a weak-field approximation defined as follows \cite{Bisnovatyi2010}
\begin{align}
    g_{\alpha \beta} =\eta_{\alpha \beta }+h_{\alpha \beta},
\end{align}
where $\eta_{\alpha \beta }$ and $h_{\alpha \beta}$ connote the Minkowski metric and perturbation metric, respectively, and their properties \cite{Bisnovatyi2010}
\begin{align}
    \eta_{\alpha \beta } = diag(-1,1,1,1),
\end{align}
\begin{align}
    \ h_{\alpha \beta} << 1, \hspace{0.5cm} h_{\alpha \beta}\to 0
    \hspace{0.5cm} \rm{under} \hspace{0.5cm} x^\alpha \to\infty
\end{align}
\begin{align}
    g^{\alpha \beta} =\eta^{\alpha \beta }-h^{\alpha \beta},\hspace{0.27cm} h^{\alpha \beta}=h_{\alpha \beta}.
\end{align}

We now want to study that plasma effects on deflection angle of the light rays.  For the presence of plasma medium, deflection angle can be expressed as \cite{Bisnovatyi2010}
\begin{align}
    \hat{\alpha}_i=\pm\frac{1}{2}\int^\infty_{-\infty}\Big[\frac{1}{2}\Big(h_{33,i}+\frac{\omega^2}{\omega^2-\omega_e^2} h_{00,i}-\frac{K_e}{\omega^2-\omega_e^2}\omega^2  N_{,i}\Big)\Big]d z.
\end{align}
$N(x^i)$ shows the number density of the particles in the plasma around the black hole and $K_{e}=4\pi e^2/m_{e}$ is a constant. The $\pm$ signs of $\hat{\alpha_i}$ determines the deflection towards and away from the central object, respectively. For large distance we can approximate the black hole metric as
\begin{align}
    d s^2=d s_0^2+\Bigg(\frac{R_s}{r}-\epsilon\frac{R_s^2}{4r^2}\Bigg)d t^2+\Bigg(\frac{R_s}{r}-\epsilon\frac{R_s^2}{4r^2}\Bigg)d r^2,
\end{align}
where $ds^2=-dt^2+dr^2+r^2(d\theta^2+\sin^2\theta d\phi^2)$ and further calculation we use $R_s=2M$ as Schwarzschild radius. In the Cartesian coordinates the components $h_{\alpha\beta}$ can be written as
\begin{align}
    h_{00}=\Big(\frac{R_s}{r}-\epsilon\frac{R_s^2}{4r^2}\Big),
\end{align}
\begin{align}
    h_{jk}=\Big(\frac{R_s}{r}-\epsilon\frac{R_s^2}{4r^2}\Big)n_j n_k,
\end{align}
\begin{align}
    h_{33}=\Big(\frac{R_s}{r}-\epsilon\frac{R_s^2}{4r^2}\Big)\cos^2{x},
\end{align}
where $\cos{x}=\frac{z}{\sqrt{b^2+z^2}}$ and $r=\sqrt{b^2+z^2}$, $b$ is the impact parameter signifying the closest approach of the photons to the black hole. Using the above mentioned expressions in the formula, one can compute the light deflection angle with respect to $b$ for a black hole  surrounded by plasma
\begin{align}\label{alpha general}
    \hat{\alpha}_b= \int^\infty_{-\infty}
 \frac{1}{2 }\Bigg[\partial_b\Bigg(\Big(\frac{R_s}{r}-\epsilon\frac{R_s^2}{4r^2}\Big)\cos^2{x}\Bigg)+\partial_b\Big(\frac{R_s}{r}-\epsilon\frac{R_s^2}{4r^2}\Big)\frac{\omega^2}{\omega^2-\omega_e^2} -\frac{K_e}{\omega^2-\omega_e^2}\partial_b N\Bigg]d z.
\end{align}
by using from $\partial_b=\frac{b}{r}\partial_r$ expression, we can get following equation

\begin{align}\label{alpha general}
    \hat{\alpha}_b= \int^\infty_{-\infty}
 \frac{b}{2 r}\Bigg[\partial_r\Bigg(\Big(\frac{R_s}{r}-\epsilon\frac{R_s^2}{4r^2}\Big)\cos^2{x}\Bigg)+\partial_r\Big(\frac{R_s}{r}-\epsilon\frac{R_s^2}{4r^2}\Big)\frac{\omega^2}{\omega^2-\omega_e^2} -\frac{K_e}{\omega^2-\omega_e^2}\partial_r N\Bigg]d z.
\end{align}
In the light of foregoing discussion, we can easily examine the impact of different plasma mediums on the photon deflection angle as depicted in Fig {\ref{alpha uniform - b }}.

\subsection{Uniform plasma}

In the first case, we consider homogeneous plasma with $\omega_0^2=\rm{const}$. In this state of plasma, the refractive index does not related with space coordinate explicitly, so we can ignore the refractive action. In other words, we do not consider last term of Eq.(\ref{alpha general}). By integrating Eq.(\ref{alpha general}), we have following result for deflection angle  
\begin{align}\label{alpha-uni}
    \hat{\alpha}_{uni}=  \Big(\frac{R_s}{b}-\epsilon\frac{\pi R_s^2}{16b^2}\Big)+\Big(\frac{ R_s}{b}-\epsilon\frac{\pi R_s^2}{8b^2}\Big)\frac{1}{1-\frac{\omega_0^2}{\omega^2}}
\end{align}

\begin{figure}
    \centering
    \includegraphics[scale=0.45]{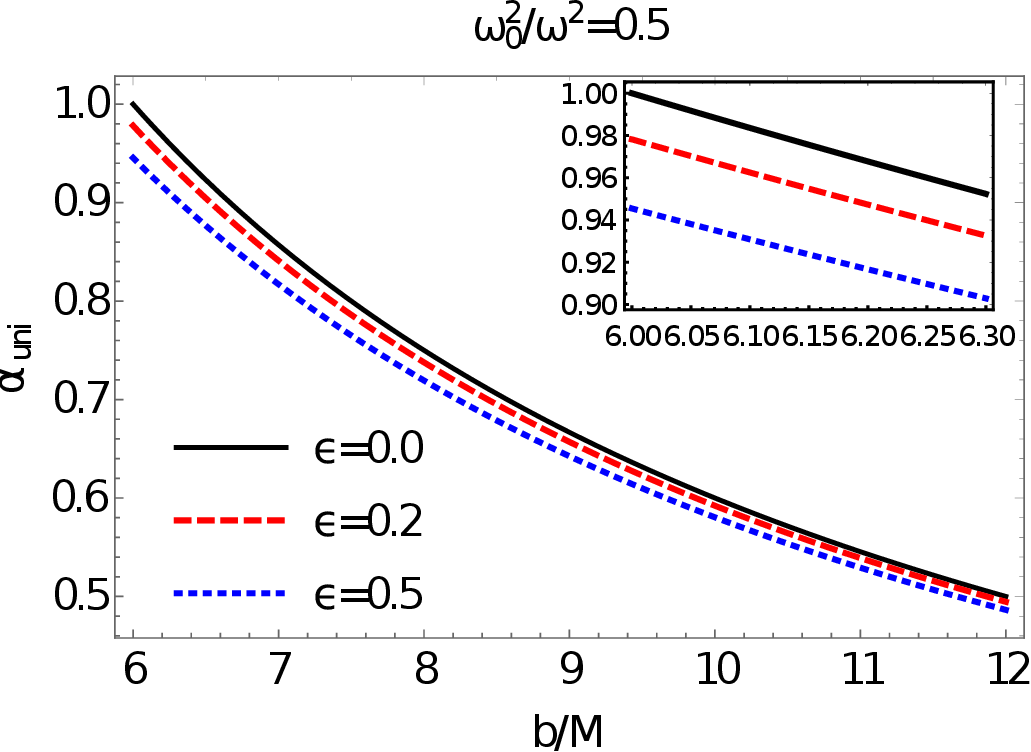},    \includegraphics[scale=0.45]{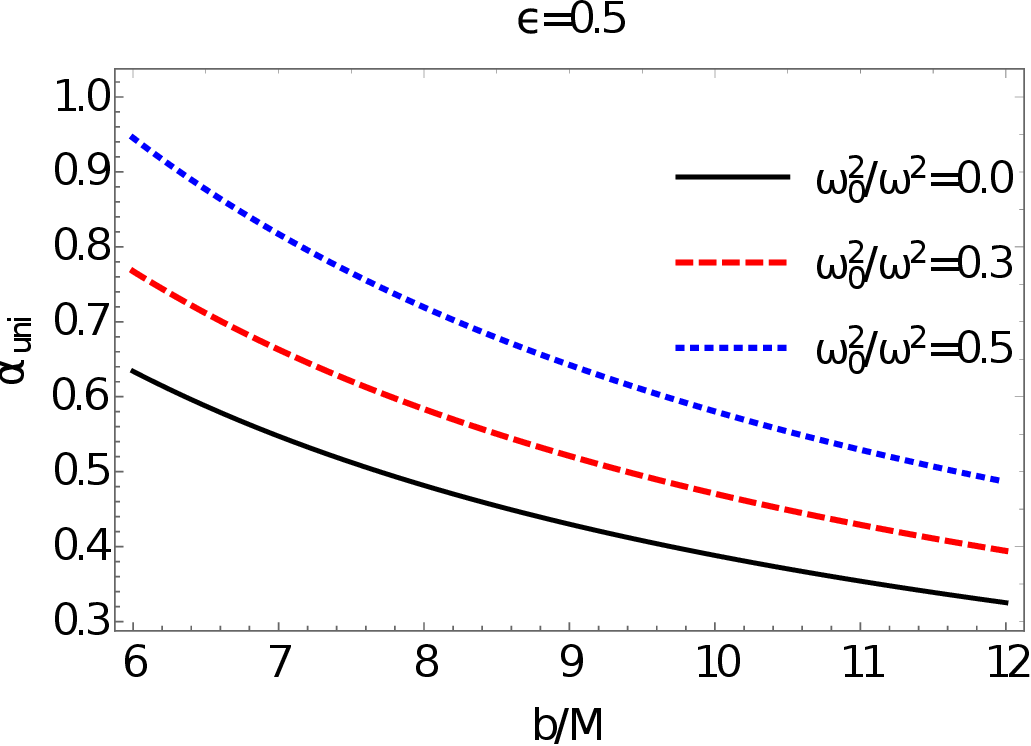}
    \caption  \ {Plot shows the deflection angle $\hat{\alpha}_{uni}$ as a function of the impact parameter b.}
    \label{alpha uniform - b }
\end{figure}

\begin{figure}
    \centering
    \includegraphics[scale=0.45]{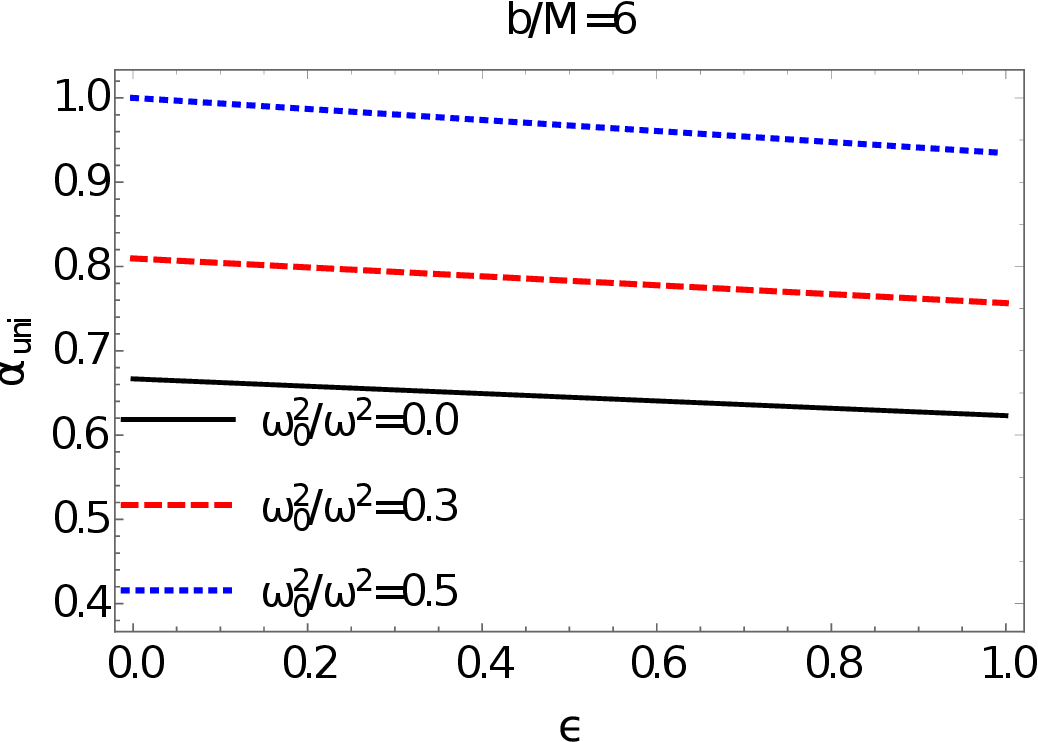}
\includegraphics[scale=0.45]{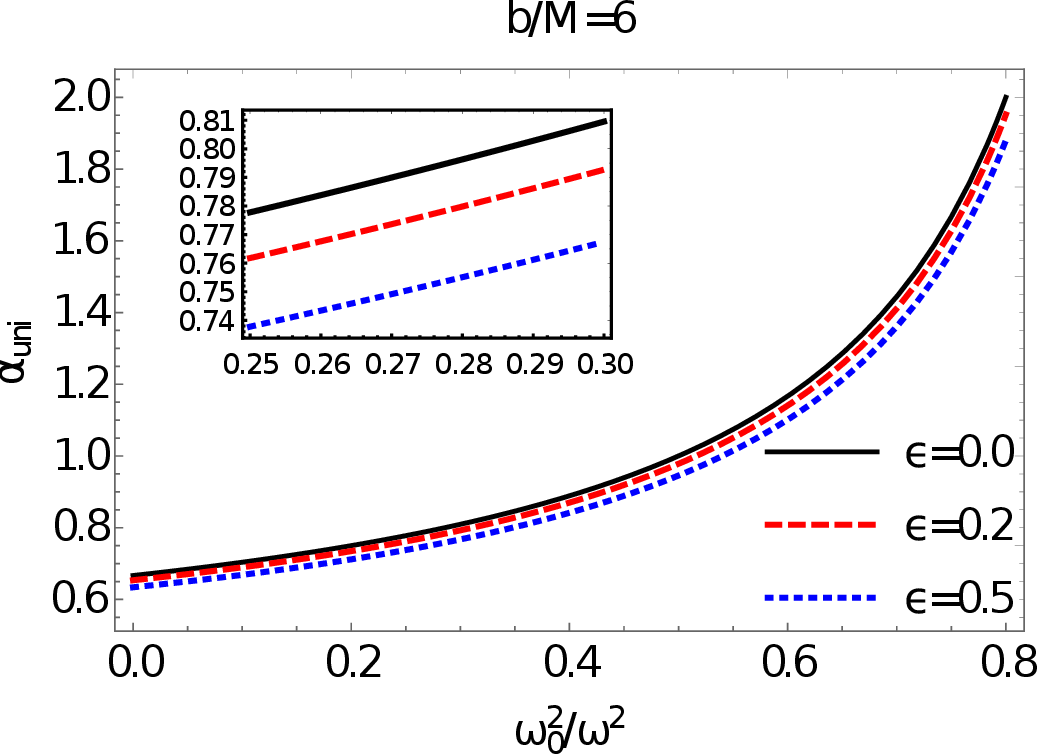}
    \caption{Plot of the deflection angle $\hat{\alpha}_{uni}$ as a function of  $\frac{\omega_0^2}{\omega^2}$  and  $\alpha$  with a fixed impact parameter $b/M=6$}
    \label{alpha uni-omega -epsilon}
\end{figure}

Plots of the impact parameter b for different GUP parameter $\epsilon$ (left panel) and plasma parameters $\frac{\omega_0^2}{\omega^2}$ (right panel) are shown in Fig.{\ref{alpha uniform - b }}.
 For decreasing of the impact parameter b, a rise in the deflection angle is investigated, which means that a masless particle moving too close to the black hole's surroundings basically increases its deviating propensity. Fig.{\ref{alpha uni-omega -epsilon}}. is a visualization of the deflection angle distinctively with respect to $\frac{\omega_0^2}{\omega^2}$ and $\epsilon$. The deflection angle is maximum due to high plasma distribution (right panel) and is seen to be strictly decreasing against an increasing GUP $\epsilon$ (right panel), for instance, taking $\epsilon=0$ the Schwarzschild gravity ensures the highest degree of deviation $\hat{\alpha}_{uni}$. We deduce that, as expected, the existence, of plasma in the black hole vicinity, contrariwise to the vacuum case $\frac{\omega_0^2}{\omega^2}=0$ contributes to the photon motion
\subsection{Singular isothermal sphere}
The Singular Isothermal Sphere (SIS) is the most suitable model for understanding the features of a gravitational lens photon. It was primarily introducted in order to explore the len's property and clusters.In general, the SIS is a spherical cloud of gas with a single feature of density up to infinity at its center. The density distribution of a SIS is given by \cite{Bisnovatyi2010,Babar21a}
\begin{align}
    \rho(r)=\frac{\sigma^2_\nu}{2 \pi r^2}
\end{align}
where $\sigma_nu^2$ refers to a one dimensional velocity. The plasma concentration admits  the following analytic dispersion. The plasma concentration admits the following analytic expression \cite{Bisnovatyi2010,Babar21a}
\begin{align}
    N(r)=\frac{\rho(r)}{\kappa m_p},
\end{align}
here $m_p$ is the proton mass and $k$ is a dimensionalless constant coefficient generally associated to the dark matter universe. Utilizing the plasma frequency takes the form 
\begin{align}
    \omega_e^2=K_e N(r)=\frac{K_e \sigma^2_\nu}{2 \pi \kappa m_p r^2 }.
\end{align}
We reckon with the above mentioned properties of the SIS and compute the angle of deflection $\hat{\alpha}_{SIS}$ as bellow
\begin{align}
    \hat{\alpha}_{SIS}=  \Big(\frac{2R_s}{b}-\epsilon\frac{3 \pi R_s^2}{16b^2}\Big)+\frac{ R_s^2 \omega_c^2}{b^2 \omega^2}\Big(\frac{1}{2}-\frac{2 R_s}{3 \pi b}+\epsilon\frac{3\pi R_s^2}{16b^2}\Big)
\end{align}
These calculations brings up a supplementary plasma constant $\omega_c^2$ which has  the following analytic expression \cite{Babar21a}
\begin{align}
    \omega_c^2=\frac{K_e \sigma^2_\nu}{2 \pi \kappa m_p R_s^2 }
\end{align}

\begin{figure}
    \centering
    \includegraphics[scale=0.35]{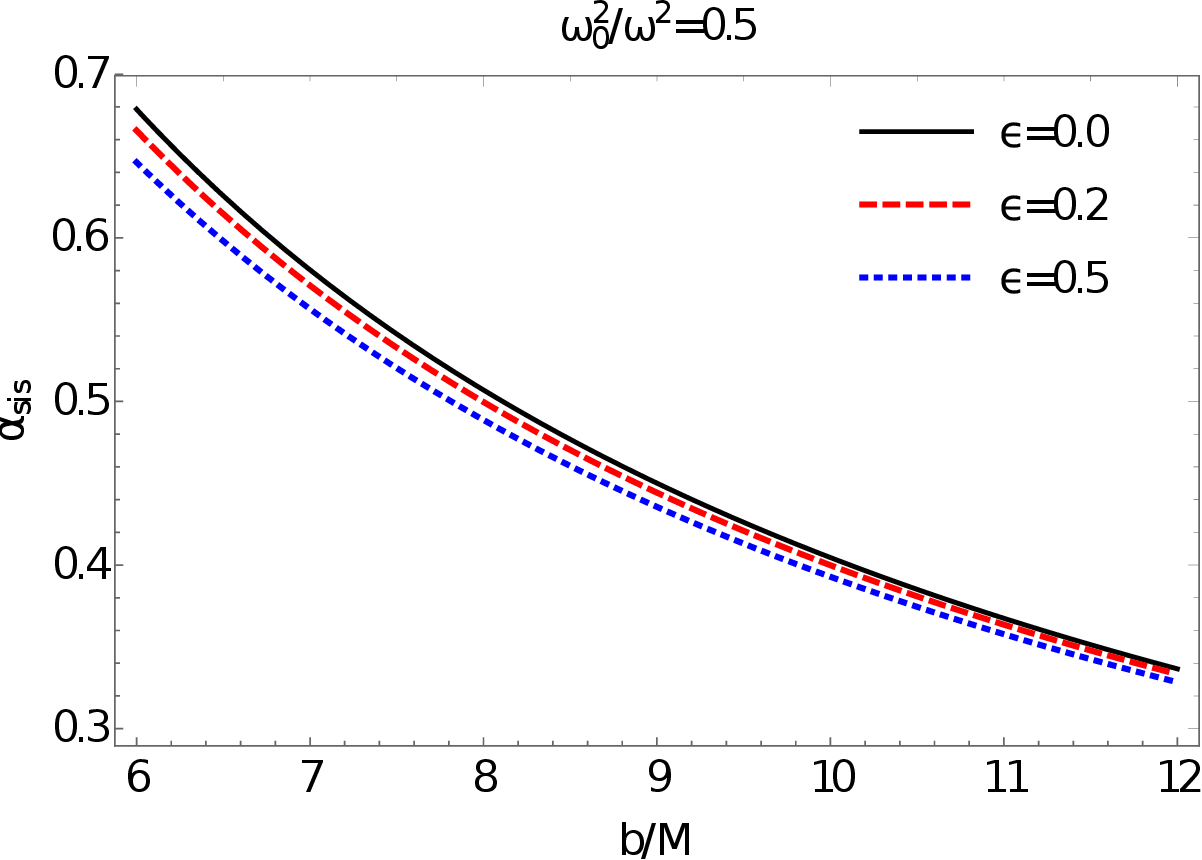}
    \includegraphics[scale=0.35]{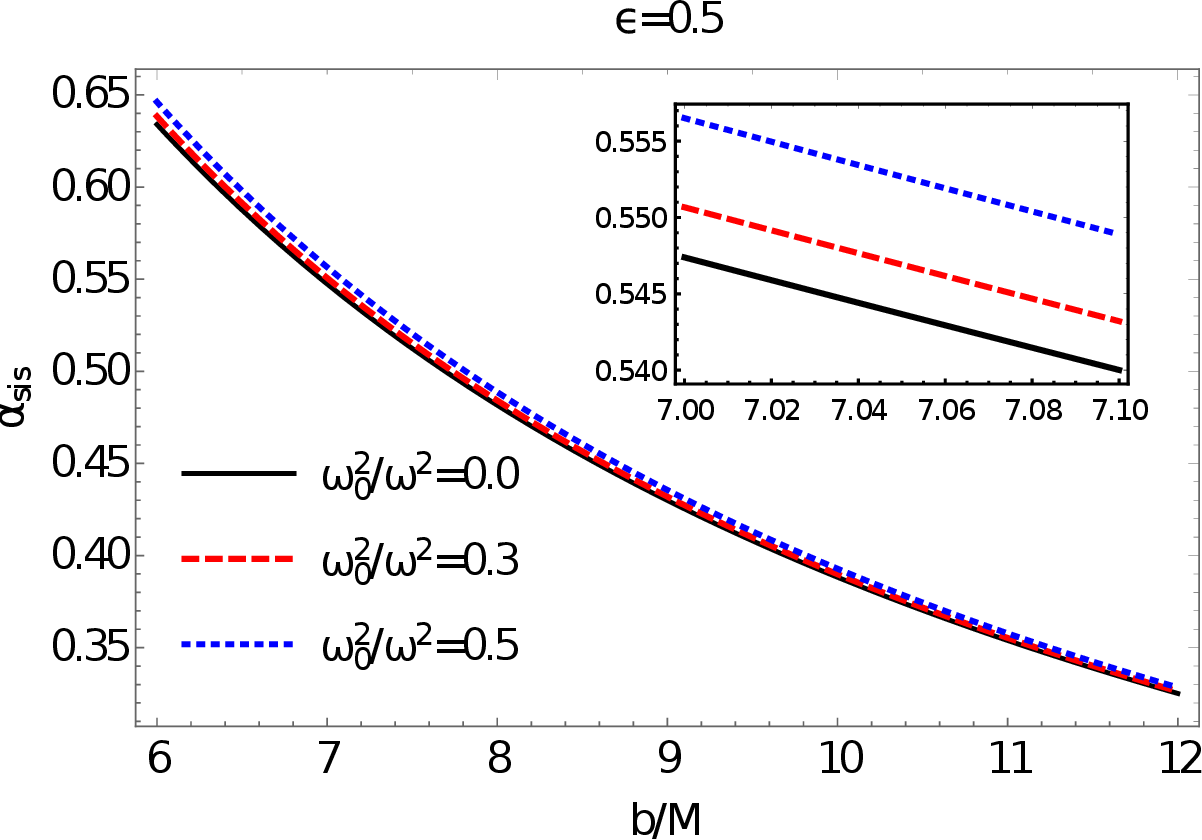}
    \caption{This graph shows a variation of $\hat{\alpha}_{sis}$ with respect to $b$ in different $\epsilon$ (left panel) and $\frac{\omega_0^2}{\omega^2}$ (right panel).}
    \label{Alpha sis -b2}
\end{figure}

\begin{figure}
    \centering
    \includegraphics[scale=0.41]{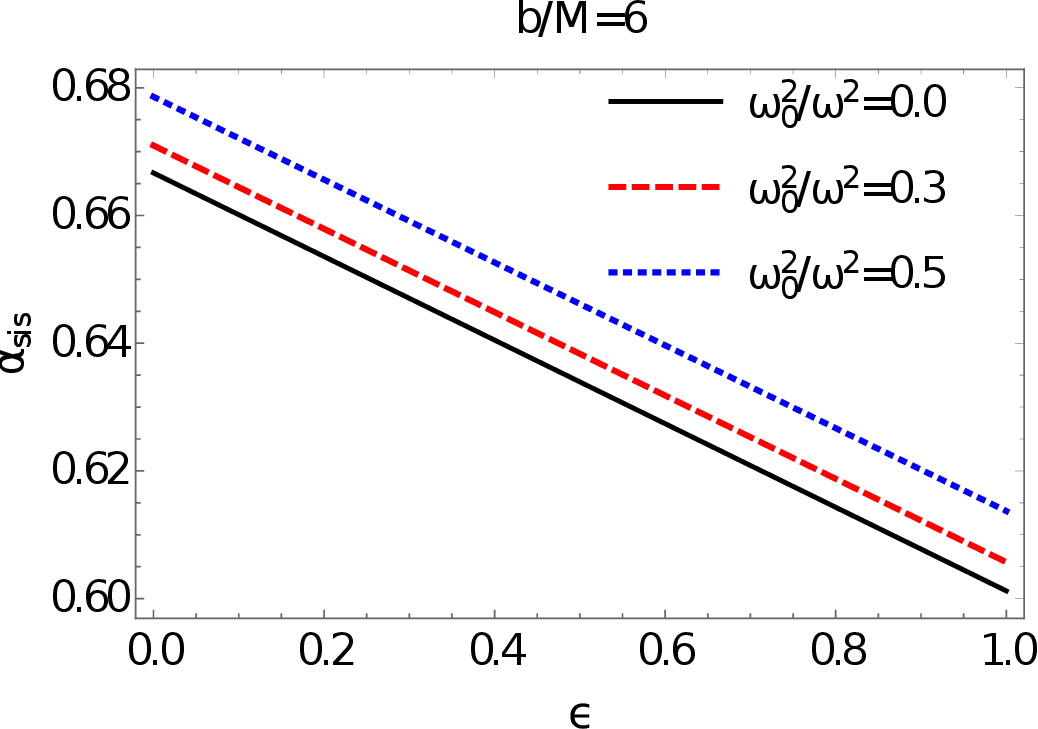} \includegraphics[scale=0.35]{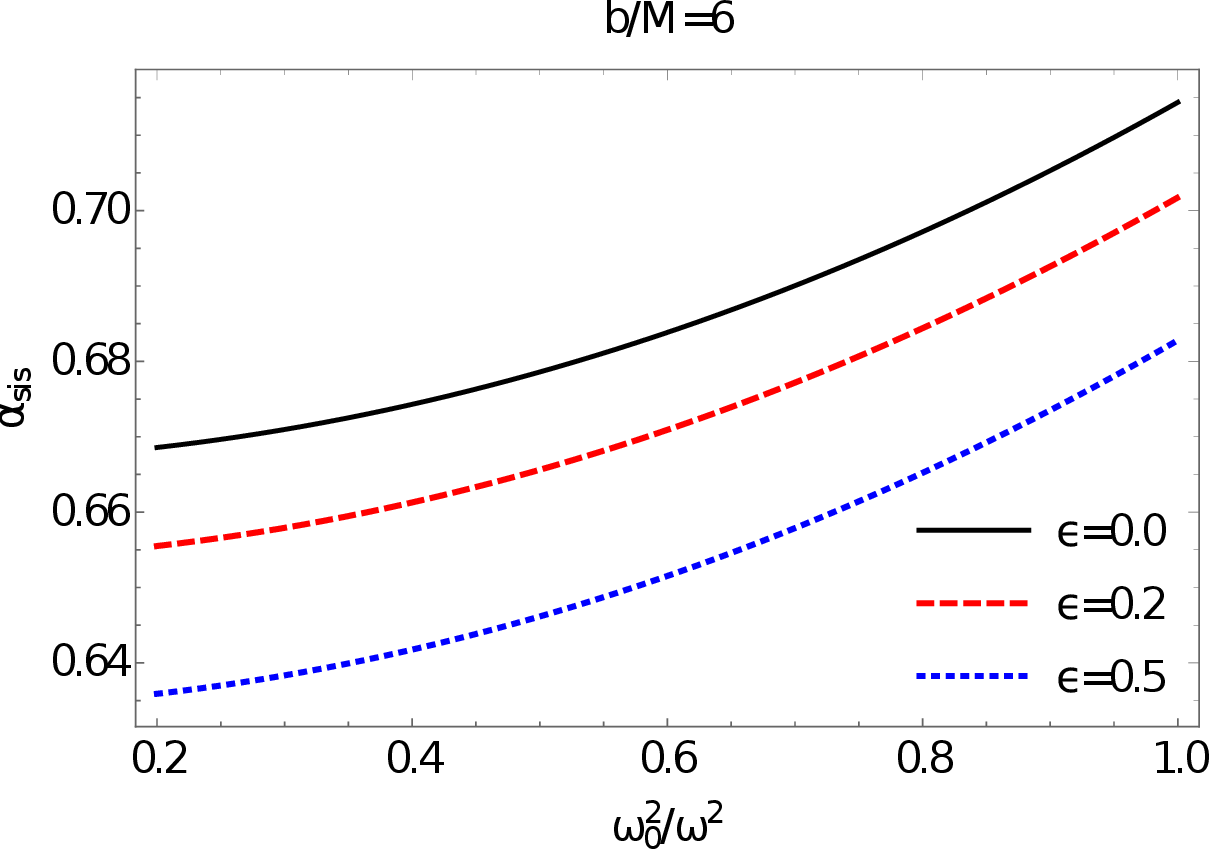}
    \caption{Plot of the deflection angle $\hat{\alpha}_{sis} $ as a function of  $\epsilon$ (  left panel ) and $\frac{\omega_0^2}{\omega^2} $ (  right panel ) for a fixed impact parameter $b/M=6$.}
    \label{alpha sis- epsilon -omega}
\end{figure}

In order to assimilate the influence of SIS on the photon trajectory we plotted the deflection angle $\hat{\alpha}_{SIS}$ as a function of the impact parameter $b$, see Fig.\ref{Alpha sis -b2}. Interestingly, we see that the uniform plasma and SIS medium share common features regarding the parameter b. Note that, the quantity $\frac{\omega_c^2}{\omega^2}$ identifies distribution of SIS in the black hole vicinity, thus we detected the photon sensitivity to specified parameter along with the coupling constant parameter $\epsilon$, by means of a graphical analysis in Fig.{\ref{alpha sis- epsilon -omega}}. We examined that $\hat{\alpha}_{SIS}$ decreases when $\epsilon$ increases (left panel) and, conversely, $\hat{\alpha}_{SIS}$ increases when $\frac{\omega_c^2}{\omega^2}$ increases (right panel) . Hence, the presence of SIS in the black hole surroundings to some extent affects the intervening massless particles.
\begin{figure}
    \centering
    \includegraphics[scale=0.35]{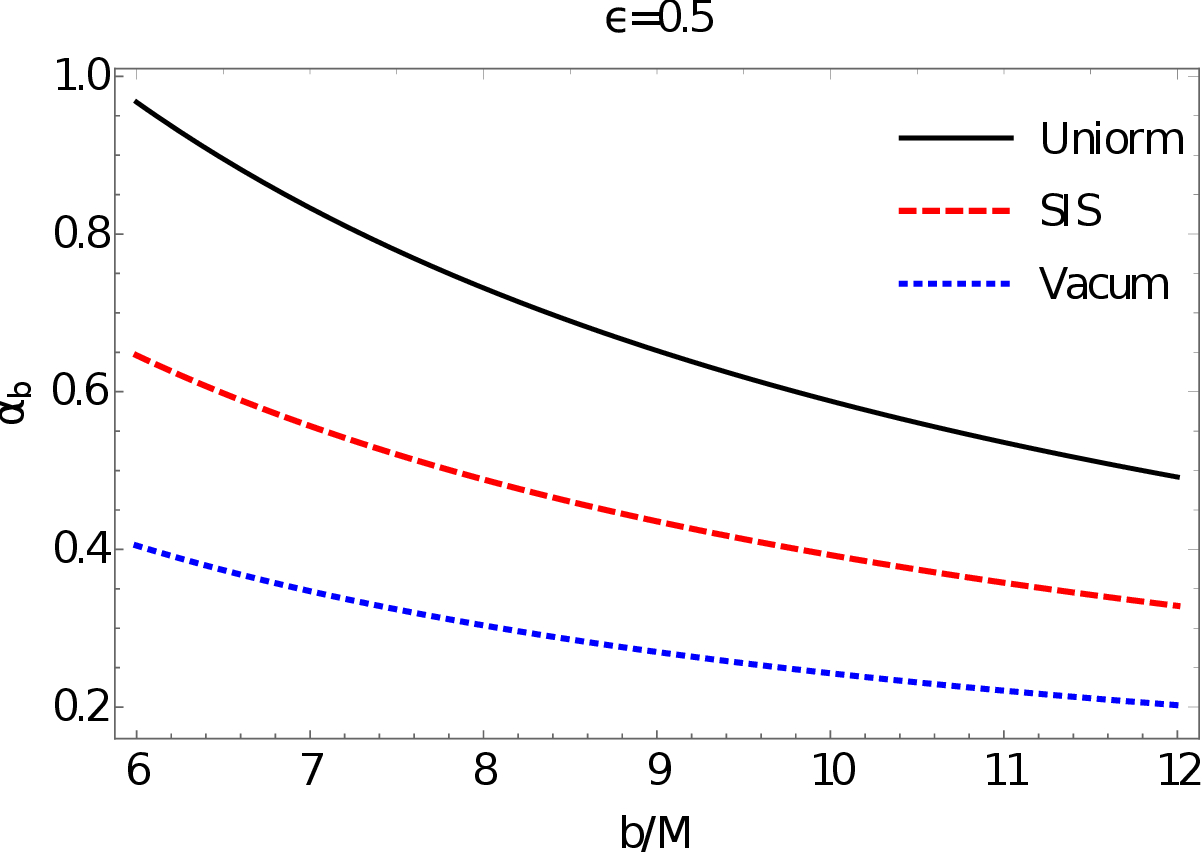}
    \includegraphics[scale=0.35]{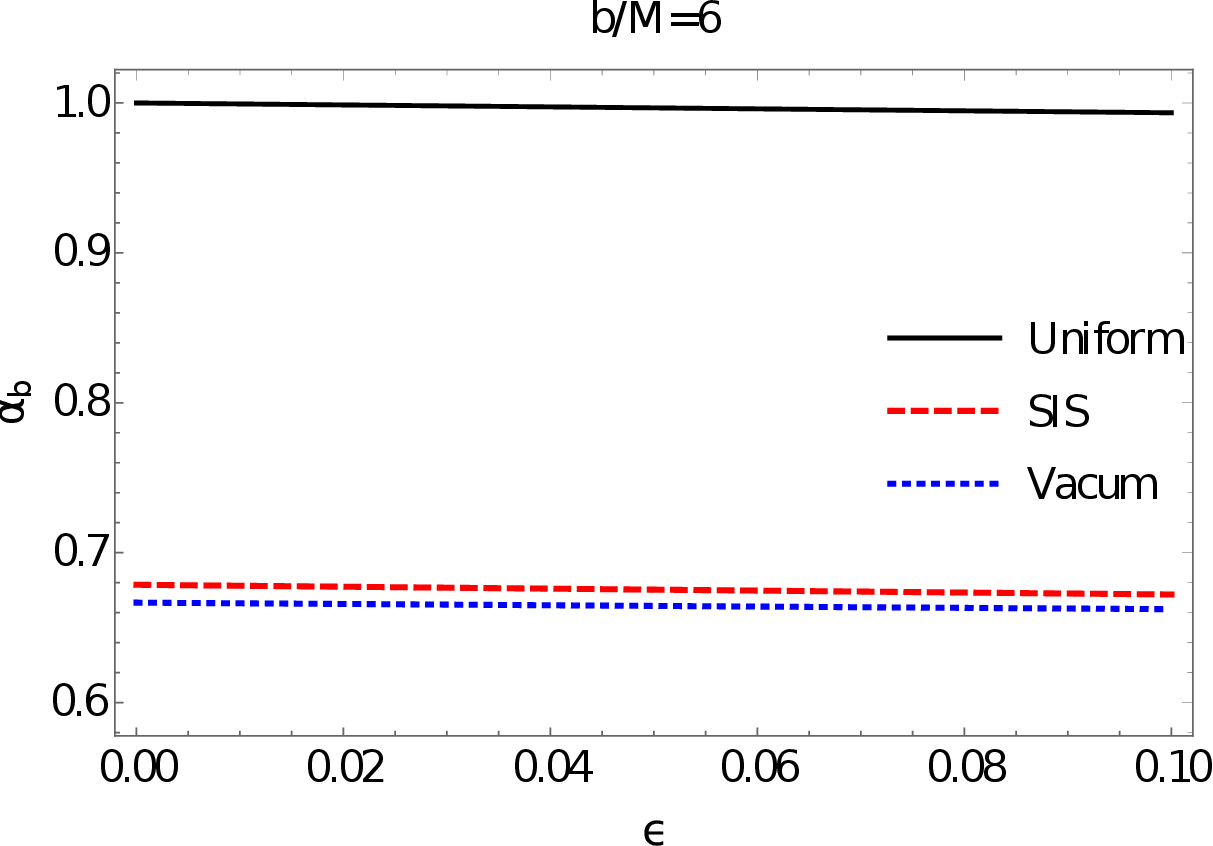}
     \caption{Plot of the deflection angle $\hat{\alpha}_{b}$ as function of the impact parameter b and $\epsilon$. The correspoding fixed parameters used are $\frac{\omega_0^2}{\omega^2}=0.5,\frac{\omega_c^2}{\omega^2}=0.5,b/M=6,r_c=3$ }
    \label{alpha-b-}
\end{figure}

Fig.{\ref{alpha-b-}} is a visual juxtaposition of the $\hat{\alpha}_{uni},\hat{\alpha}_{SIS}$ as a function of the impact parameter and the parameter $\epsilon$. It is quite obvious that the deflection is maximum when the black hole is surrounded by a uniform plasma medium. The final result can therefore be encapsulated in a mathematical expression as, $\hat{\alpha}_{uni}>\hat{\alpha}_{SIS}$

\begin{figure}
    \centering
    \includegraphics[scale=0.4]{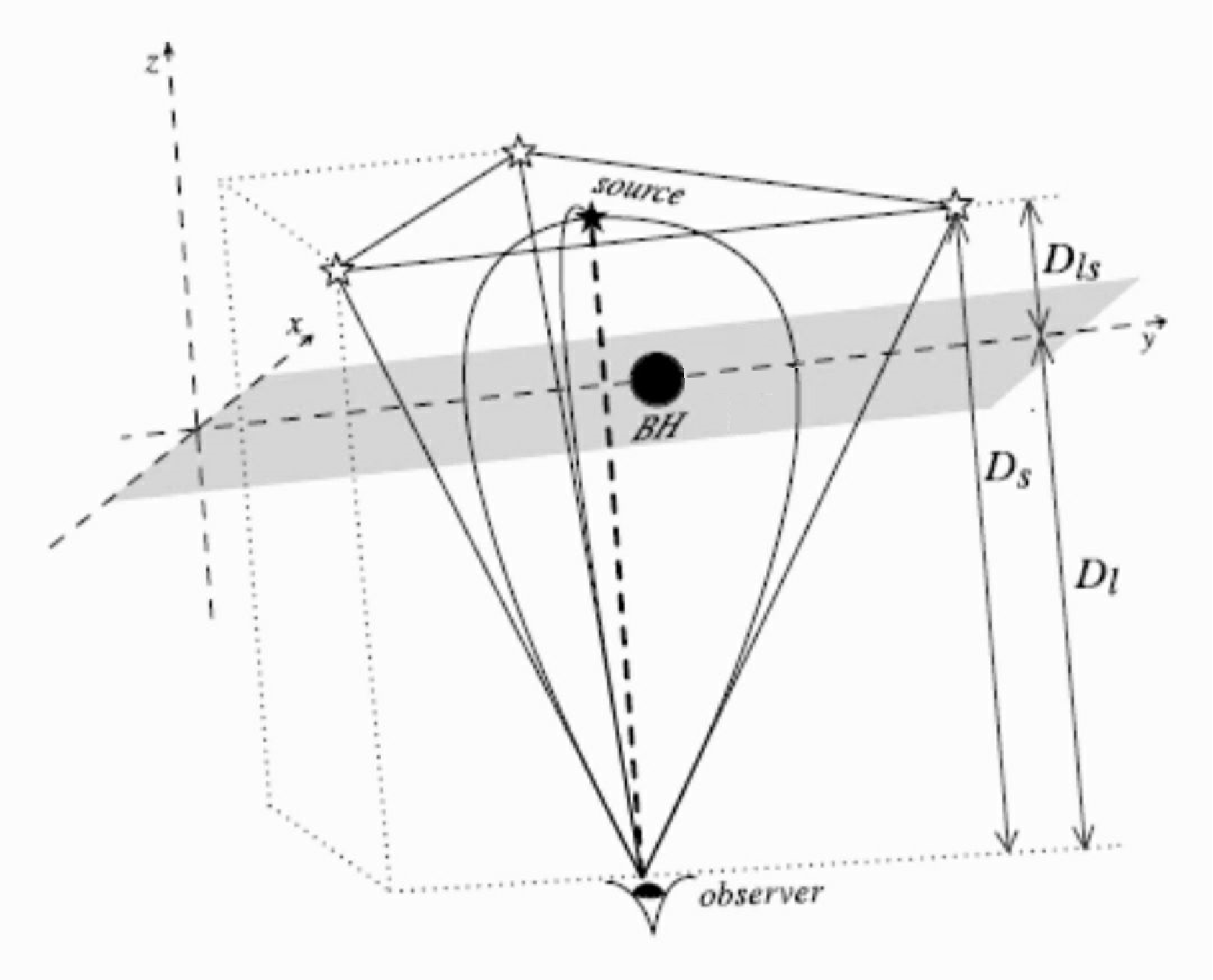}
    \caption{Schematic wiew of gravitational lensing adapted from \cite{Morozova2013Ap&SS}.}
    \label{schema}
\end{figure}

\subsection{Lens equation and magnification}

 We now focus on observable consequences of the gravitational lensing namely the magnification of the image source brightness in the presence of plasma performing the angle of deflection $\hat{\alpha}$ discussed  in our previous sections, particularly for the uniform plasma. The schematic diagram Fig.(\ref{schema}) depicts the gravitational lensing system showing the black hole as lens, source and the observer. For finding magnification of the image sources, we can use lens equation, which is related to angular position of the source $\beta$ and image $\theta$ and distances from the source to the observer $D_s$ and to the lens $D_{ls}$ and deflection angle \cite{Virbhadra:1999nm,Virbhadra:2008ws,Bozza:2010xqn,Younas:2015sva,Azreg-Ainou:2017obt}:
\begin{align}
    \theta D_s= \beta D_s + \hat{\alpha} D_{ls}.
\end{align} 
In the weak field limit, we can use $b\approx D_d \theta$ relation. Then we have following equation
    \begin{align}\label{beta}
        \beta=\theta-\frac{D_{ds}}{D_s }\hat{\alpha}.
    \end{align}
In the case of uniform plasma, by using Eq.(\ref{alpha-uni}) for $\hat{\alpha}$ we can rewrite Eq.(\ref{beta}) as following form 
    \begin{align}\label{magnification}
        \beta=\theta-\frac{D_{ds}}{D_s D_d}(1+\frac{1}{1-\frac{\omega_0^2}{\omega^2}})\frac{R_s}{\theta}+\frac{D_{ds}}{D_s D_d^2}(1+\frac{2}{1-\frac{\omega_0^2}{\omega^2}})\frac{\epsilon\pi R_s^2}{16\theta^2}.
    \end{align}
By introducing new variable $x=\theta-\frac{\beta}{3}$, we can reduce Eq.(\ref{magnification}) to the form
\begin{align}\label{equation}
    &x^3+px+q=0, \,
\end{align}
where 
\begin{align}
    p&=-\frac{\beta^2}{3}-\frac{1}{2}(1+\frac{1}{1-\frac{\omega_0^2}{\omega^2}})\theta_E^2, \\
    q&=-2\frac{\beta^3}{27}-\frac{\beta}{6}(1+\frac{1}{1-\frac{\omega_0^2}{\omega^2}})\theta_E^2+\frac{\epsilon\pi R_s}{32D_d}(1+\frac{2}{1-\frac{\omega_0^2}{\omega^2}})\theta_E^2 \\
    \theta_E&=\sqrt{\frac{2R_s D_{d s}}{D_s D_d}}\ .
\end{align}
Hence the solution of Eq.(\ref{equation}) is given by \cite{Turi:2019a}
\begin{align}
    x=2s^{1/3}\cos{\frac{\phi+2k\pi}{3}},\,
\end{align}
where 
\begin{align}
    s=\sqrt{-\frac{p^3}{27}}~, \,\,\, \phi=\arccos{(-\frac{q}{2s})}.
\end{align}
The total magnification of the images is defined as the ratio of solid angles of the observer to the source and than summed over relative to each image. However in weak field approximation, the total magnification $\mu_{\Sigma}$ of the images is approximated as \cite{Morozova2013Ap&SS}
    \begin{align}
     \mu_{\Sigma}=\frac{I_\mathrm{tot}}{I_*}=\underset{k}\sum\bigg|\bigg(\frac{\theta_k}{\beta}\bigg)\bigg(\frac{d\theta_k}{d\beta}\bigg)\bigg|, \quad k=1,2,...,n
    \end{align}
where $\theta_k=x+\beta/3$. There are three components of the magnification of the images which is ordinary two images and relativistic images appeared due to presence of $\epsilon$ parameter.
      \begin{figure}
          \centering
          \includegraphics[width=0.46\textwidth]{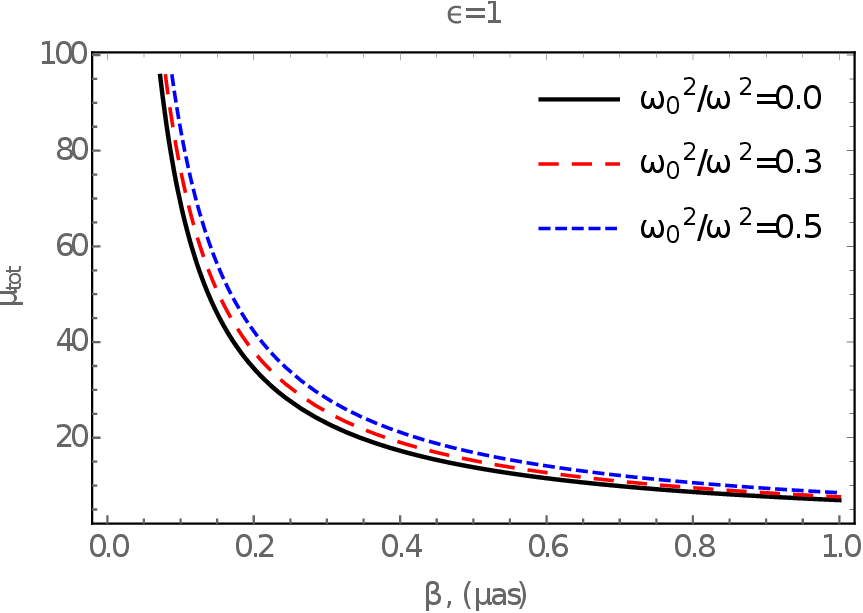}
          \includegraphics[width=0.45\textwidth]{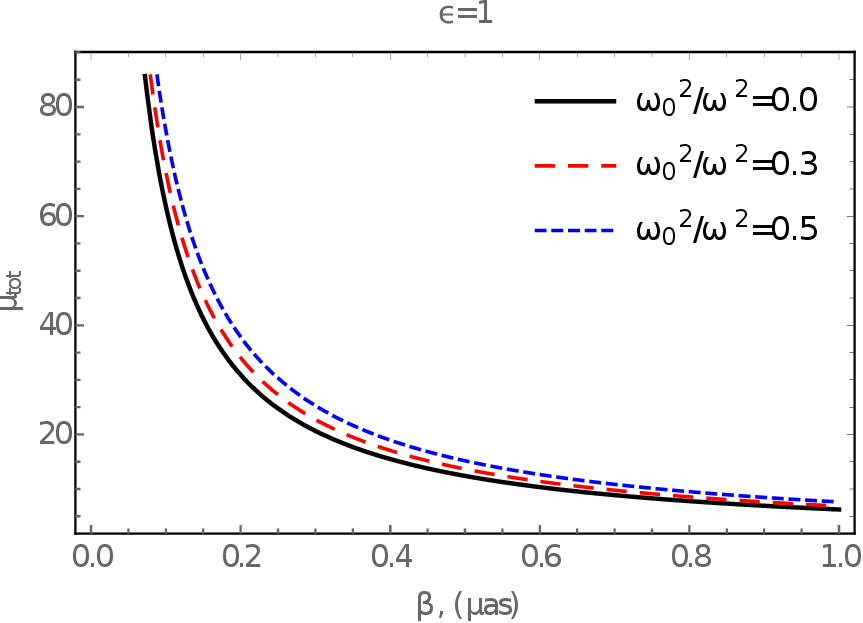}
          \caption{Total magnification of the image source as a function of position of source $\beta$ for $\epsilon=1$, $\omega_0^2/\omega^2=0.5$ and $M/D_d=2.38\cdot 10^{-11}$(left panel) as well as $M/D_d=1.911\cdot 10^{-11}$(right panel).}
          \label{magnification-b}
      \end{figure}

      \begin{figure}
          \centering
          \includegraphics[width=0.6\textwidth]{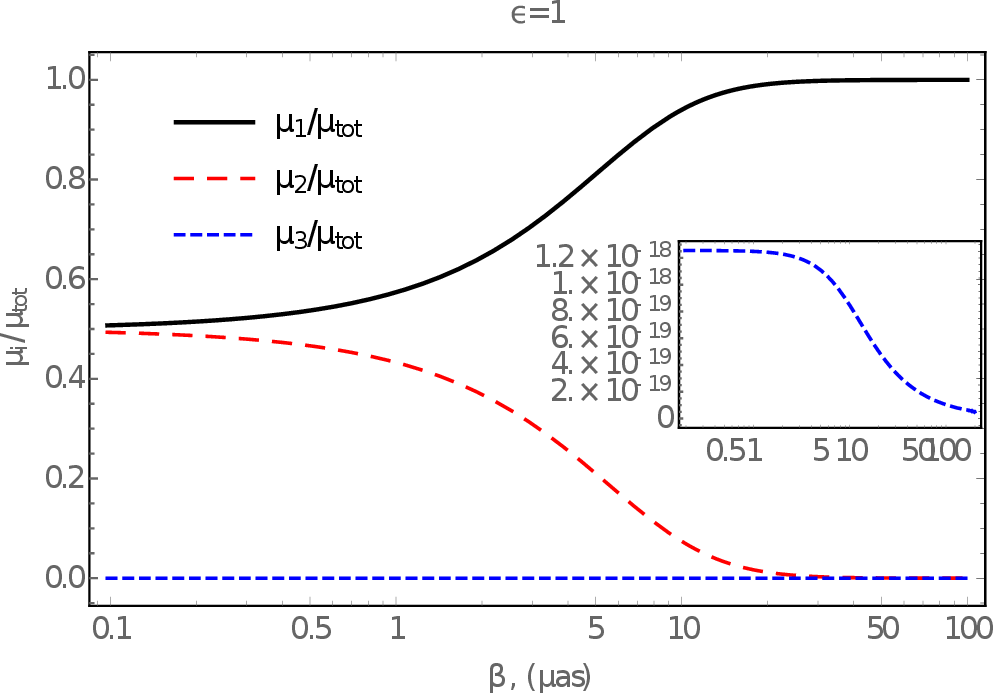}
          \caption{Magnification rate of the image source as a function of position of source $\beta$ for $\epsilon=1$, $\omega_0^2/\omega^2=0.5$ and $M/D_d=2.26\cdot 10^{-11}$. Black thick and red dashed curves represents magnification rate of primary and secondary images, respectively, while red curve related to relativistic images due to presence of $\epsilon$ parameter.}
          \label{magnification-r}
      \end{figure}

Using above expressions we can get numerical results for total magnification. Fig.(\ref{magnification-b}) illustrates dependence of total magnification of the image source on source angular position $\beta$ for different value of $M/D_d$ quantities. These quantities have been taken from \cite{2} (for left panel) and \cite{Akiyama2022sgr} (for right panel). From these pictures one can see that total magnification decreases with increase of $\beta$ position. Also, it is possible to see that the presence of the plasma medium influence to grow of the magnification. 

 \begin{table*}[t]
\caption{Comparison of magnification factor of images in homogeneous plasma and vacuum. We have used in this table that $M/D_d=2.26\cdot10^{-11}$ what corresponds to supermassive black hole in center of Milky Way, $\beta=1\, \mu as$ and $D_s/D_{ds}=2$(these parameters have taken from \cite{Virbhadra:2008ws}).}
\begin{tabular}{|c|c|c|c|c|}
\hline\hline
 $\mu_n/\mu_n{}^{\text{vac}}$ & $\text{$\omega_0^2 $/}\omega^2\text{=0.1}$ & $\text{$\omega_0^2 $/}\omega^2\text{=0.5}$ & $\text{$\omega_0^2 $/}\omega^2\text{=0.9}$ & $\mu _n^{\text{vac}}$ \\
 \hline
 $\mu_1/\mu _1{}^{\text{vac}}$ & 1.0235 & 1.19297 & 2.15873 & 3.89282 \\
 $\mu _2/\mu _2{}^{\text{vac}}$ & 1.03162 & 1.25967 & 2.55929 & -2.89282 \\
 $\mu_3/\mu_3{}^{\text{vac}}$ & 0.947295 & 0.666667 & 0.181862 & $-1.55301\cdot 10^{-17}$ \\
 \hline\hline
\end{tabular} \label{table1}
\end{table*}

\begin{table*}[t]
\caption{Approxmated value of angular position of images, deflection angles and magnifications. All of angular quantities given by $\mu as$ and $1,2,3$ indexes means that these quantities belong to primary, secondary and third images. All calculation is done with $\beta=1 (\mu as)$, $\epsilon=1$ and $\omega_0^2/\omega^2=0.5$.}
    \begin{tabular}{|c|c|c|c|c|c|c|c|c|c|}
    \hline\hline
 \text{SMBH} & $\theta_1$ & $\alpha _1$ & $\mu _1$ & $\theta _2$ & $\alpha _2$ & $\mu _2$ & $\theta _3$ & $\alpha _3$ & $\mu _3$ \\
 \hline 
 \text{$M87^*$} & 8.088 & $17\cdot10^{-5}$ & 4.310 & -7.088 & $-1.617\cdot10^{-5}$ & -3.310 & $7.504\cdot10^{-6}$ & $-2.011\cdot10^{-6}$ & $-7.371\cdot10^{-18}$ \\
 \text{$Sgr A^*$} & 8.964 & $1.592\cdot10^{-5}$ & 4.747 & -7.964 & $-1.792\cdot10^{-5}$ & -3.747 & $9.346\cdot10^{-6}$ & $-1.999\cdot10^{-6}$ & $-1.143\cdot10^{-17}$ \\
 \hline\hline
\end{tabular}\label{table2}
\end{table*}
Comparison magnification of images in homogeneous plasma and in vacuum represents in Table~\ref{table1}. One can see that magnification of third image decrease in the presence of plasma. On the contrary, for primary and secondary images magnification increase with decreasing of plasma frequencies. In addition, we can find values of position of image, deflection angle and magnification of the images for different SMBH. Table~\ref{table2} depicts value of these quantities for M87* and Sgr A*. 

In fig.(\ref{magnification-r}), it is shown that relation of the magnification ratio $\mu_i/\mu_{tot}$ and position of the source $\beta$. We can see that in large value of $\beta$ magnification of primary image will be main term of total magnification. On other word, magnification of secondary and third images tend to zero in the great value of $\beta$. 

\section{Conclusion}\label{conlusion}

In this paper, we have tested Schwarzschild spacetime with modified GUP parameter through studying null geodesic, shadow and gravitational weak lensing.
From above investigating, we can summarize as follows:
\begin{itemize}

\item Studying of the null geodesic we can observe that photon orbits decrease due to presence of the parameter $\epsilon$. As well as Radius of the BHs shadow decrease with increasing of the parameter $\epsilon$.

\item During studying weak gravitational lensing, one can observe that the presence of GUP modification parameter $\epsilon$ causes the deviation angle to decrease. 

\item In Figs.(\ref{magnification-b},\ref{magnification-r}) we can conclude what due to the presence of plasma due to the presence of the parameter, BH has three images, ordinary two images and a relativistic images. Magnification for these picture is shown in Fig.(\ref{magnification}).

\item Also, we have obtained influence of plasma in photon radius, shadow, deflection angle and magnification. From Figs.(\ref{photon1}) and (\ref{fig2}), photon and shadow radius decreasing with increasing plasma parameter in the case $q=1$ and for the case $q=3$ photon radius falls when plasma parameter growth. Furthermore, deflection angle and magnification of the images increase in the case of the presence of plasma. 

\end{itemize}

\section*{Acknowledgements} This research is partly supported by Research Grants FZ-20200929344 and F-FA-2021-510 of the Uzbekistan Ministry for Innovative Development. 

 

\bibliographystyle{apsrev4-1}  
\bibliography{paper,gl,LE}

\end{document}